\documentclass[12pt]{article}
\usepackage{epsfig}
\usepackage{a4,isolatin1}
\usepackage{amsmath,amsfonts,latexsym, amssymb}
\newtheorem{satz}{Theorem}[section]
\newtheorem{defi}[satz]{Definition}

\newtheorem{bem}[satz]{Remark}
\newtheorem{lemma}[satz]{Lemma}
\newtheorem{koro}[satz]{Corollary}
\newtheorem{bsp}[satz]{Example}

\newtheorem{conclusion}[satz]{Conclusion}
\newtheorem{ob}[satz]{Observation}

\newtheorem{prop}[satz]{Proposition}

\newcommand{\mcal}{\mathcal}

\newcommand{\tit}{\textit}

\newcommand{\R}{\mathbb{R}}

\newcommand{\om}{\omega}
\newcommand{\Om}{\Omega}

\begin{document}
\thispagestyle{empty}
\begin{center}
\vspace*{1.0cm}

{\LARGE{\bf Fluctuation Operators and Spontaneous Symmetry Breaking}}

\vskip 1.5cm

{\large {\bf Manfred Requardt }}

\vskip 0.5 cm

Institut f\"ur Theoretische Physik \\
Universit\"at G\"ottingen \\
Bunsenstrasse 9 \\
37073 G\"ottingen \quad Germany\\
(E-mail: requardt@theorie.physik.uni-goettingen.de)

\end{center}

\vspace{1 cm}

\begin{abstract}
  We develop an alternative approach to this field, which was to a
  large extent developed by Verbeure et al. It is meant to complement
  their approach, which is largely based on a non-commutative central
  limit theorem and coordinate space estimates. In contrast to that we
  deal directly with the limits of $l$-point truncated correlation
  functions and show that they typically vanish for $l\geq 3$ provided
  that the respective scaling exponents of the fluctuation observables
  are appropriately chosen. This direct approach is greatly simplified
  by the introduction of a smooth version of spatial averaging, which
  has a much nicer scaling behavior and the systematic developement of
  Fourier space and energy-momentum spectral methods. We both analyze
  the regime of normal fluctuations, the various regimes of poor
  clustering and the case of spontaneous symmetry breaking or
  Goldstone phenomenon.
\end{abstract} \newpage
\setcounter{page}{1}
\section{Introduction}
In the past decade in a series of papers Verbeure and coworkers
developed a beautiful and ingeneous framework to
study so-called macroscopic fluctuation phenomena in systems and
various regimes of quantum statistical mechanics (see the cited
literature). The approach is to a large extent based on a quantum
variant of the \tit{central limit theorem} and is mainly performed in
real (i.e. configuration) space. Among other things, the general goal
is it, to study the limit behavior of correlation functions of
so-called \tit{fluctuation observables}, i.e. appropriately
renormalized averages of microscopic observables, averaged over
volumes, $V$, which approach the whole space, $\R^n$, say. Typically,
one arrives, depending on the type of clustering of the microscopic
$l$-point functions, at certain simple limit algebras as e.g. $CCR$.

We approach the field from a slightly different angle. In a first step
we choose another averaging procedure, which avoids sharp volume
cut-offs and, a fortiori, has a very nice and transparent scaling
behavior. This is then exploited in the following analysis which
systematically develops so-called Fourier-space and energy-momentum
spectral methods of observables and correlation functions. We consider
it to be an advantage that the calculations turn out to be relatively
transparent and lead in a direct way to the desired results.

We first treat the case of \tit{normal fluctuations} and
$L^1$-clustering. We show that all the truncated $l$-point functions
vanish for $l\geq 3$ while they approach a finite, non-trivial limit
for $l=2$. The analysis is done both for the $(k=0)$- and the $(k\neq
0)$-modes. We emphasize that the calculations for net-momentum
different from zero remain also very simple. A variant of the method
is then applied to the case of $L^2$-clustering.

In the second part of the paper we embark on the analysis of
fluctuations in the presence of \tit{spontaneous symmetry breaking
  (ssb)}. In a first step we prove some general results in the context
of $ssb$ and the \tit{Goldstone phenomenon}. We then address the
problem of macroscopic fluctuations within this context. Among other
things, we give a general and rigorous proof that the limit
fluctuations are always classical for temperature states (a phenomenon
already observed by Verbeure et al in various simple models). The paper
ends with a treatment of extremely poor clustering, which can be
controlled by a new method we develop in the last section.
To sum up, we think that in our view the two different frameworks seem
to neatly complement each other and should lead to further interesting
results if being combined.
\section{The Scenario of Normal Fluctuations}
The following analysis works for statistical equilibrium states and/or
for vacuum states in quantum field theory. To avoid constant
mentioning of the respective scenario we are actually working in, we
usually treat equilibrium (i.e. KMS-) states, to fix the framework. Now, let
$\Omega$ be the vacuum or equilibrium state (rather its
GNS-representation; usually we work within a concrete Hilbert
space, $\mcal{H}$). As an abstract state we denote it by $\omega$. Expectations of
observables are written as
\begin{equation}\langle A\rangle=
  \omega(A)=(\Omega,A\Omega)\end{equation}
with $A$ taken from the \tit{local algebra}, $\mcal{A}_0\subset
\mcal{A}$, the latter one being the \tit{quasi-local} norm closure of
$\mcal{A}_0$. We assume $\Omega$ to be \tit{cyclic} with respect to
$\mcal{A}_0$ or $\mcal{A}$. That is, we assume
\begin{equation}\overline{\mcal{A}_0\cdot\Omega}=\mcal{H}\end{equation}
There are certain differences as to the (assumed) locality properties
of the dynamics between (non-)relativistic statistical mechanics and
relativistic quantum field theory (RQFT). Denoting the time evolution
(acting on the algebra of observables) by $\alpha_t$, we are
confronted with the following phenomenon.
\begin{ob}In RQFT part of the usual framework is the assumption
\begin{equation}\alpha_t:\;\mcal{A}_0\to \mcal{A}_0\end{equation}
while in statistical mechanics (due to weaker locality behavior) we
have in the generic case only
\begin{equation} \alpha_t:\;\mcal{A}\to\mcal{A}\end{equation}
whith $\mcal{A}_0$ usually not left invariant as the
observables will typically develop infinitely extended tails.
\end{ob}
Furthermore, we assume once for all that our system is in a \tit{pure,
  translation invariant phase}, that is $\Omega$ is extremal
translation invariant under the space translations ( which can, as in
the case of lattice systems, also be a discrete subgroup).  There can
of course exist several coexisting pure phases at the same external
parameters as in the regime below a phase transition threshold. These
assumptions imply that we can expect certain \tit{cluster properties},
i.e. decay of \tit{correlations} (see e.g. \cite{Ruelle}).
\subsection{Definition of Ordinary Fluctuation Operators}
We begin by defining the \tit{fluctuation operators} in the
\tit{normal} situation as it was done in \cite{Verbeure1}.

We assume, for the time being, $L^1$-clustering for the
two-point-function, that is
\begin{equation}\int|\langle A(x)B\rangle^T|d^nx<\infty\;A,B\in\mcal{A}\end{equation}
with $A(x)$ the translate of $A$ and
\begin{equation}\langle AB\rangle^T=\langle AB\rangle-\langle
  A\rangle\cdot\langle B\rangle\end{equation}
Once for all we assume, to simplify notation, in our particular
context that the occurring observables
are normalized to $\langle A\rangle=0$ unless otherwise stated.
\begin{defi}We define the normal (finite volume) fluctuation operators as
\begin{equation}A_V^F:=1/V^{1/2}\cdot\int_V A(x)d^nx=:1/V^{1/2}\cdot
  A_V\end{equation}
\end{defi}
In a next step one wants to give sense to these objects in the limit
$V\to\infty$. From the $L^1$-condition we however infer
\begin{equation}|(A_V^F\Omega,B\Omega)|\leq 1/V^{1/2}\int_{\R^n}
  |(A(x)\Omega,B\Omega)|d^nx\to 0\end{equation}
Hence $A_V^F\Omega\to 0$ on a dense set. Furthermore we have
\begin{multline}(A_V^F\Omega,A_V^F\Omega)=
1/V\int_V\int_V(A(x)\Omega,A(y)\Omega)dxdy=\\
1/V\int_V
  dx\left (\int_{V-x}\langle A^*A(y-x)\rangle d(y-x)\right
  )\end{multline}
This is less or equal to
\begin{equation}\label{fluc}  (1/V)\cdot V\cdot \sup_x(\int_{V-x}|(\ldots)|)\leq
\int_{\R^n}|F(y-x)|d^n(y-x)<\infty\end{equation}
(for convenience we sometimes denote a general two-point
function by $F(x-y)$). This suffices to prove weak convergence to zero for
$A_V^F\Omega$ on
the total Hilbert space $\mcal{H}$.

\begin{bem}We note that this proves also the well-known
  {\em normal-fluctuation} result $\langle A_V\cdot A_V\rangle\lesssim
  V$ in the $L^1$-case.  Under certain well-specified conditions the
  fluctuations can even be weaker than {\em normal}. If e.g. $Q_V$ is
  the local integral over a conserved quantity we proved a divergence
  significantly weaker than $\sim V$ (cf. \cite{Requ1}). But in
  general the local fluctuations will diverge in the limit
  $V\to\infty$ in contrast perhaps to ordinary intuition, even if the
  quantity is {\em globally conserved} due to quantum fluctuations
  (see also the section about {\em spontaneous symmetry breaking})
\end{bem}
A weaker than normal divergence can occur in the following situation. An asymptotic behavior $\sim
V$ does only prevail if $\int_V F(u)du\neq 0$ in the limit $V\to\infty$. On the other
side such correlation functions tend to oscillate about zero (for physical
reasons; there are e.g. usually preferred relative positions in, say, a quantum
liquid). In other words, while
\begin{equation}\int F(u)du=0\end{equation}
may seem to be rather ungeneric at first glance, it can nevertheless
happen in a specific context. The general situation is analyzed in the above reference;
certain examples of better than normal fluctuations were also found by
Verbeure et al in e.g. \cite{verbeure3} (see also \cite{Aizenman}).

For the fluctuation operators themselves we have due to
locality for $A,B\in\mcal{A}_0$:
\begin{equation}[A_V,B]\quad\text{independent of $V$ for $V\supset
    V_0\supset V_B$}\end{equation}
for some $V_0$ which contains the localisation region $V_B$ for $B\in
\mcal{A}_0$. We then have
\begin{equation}\lim_V(A_V^F\cdot C\Om,B\Om)=\lim_V([A_V^F,C]\Om,B\Om)+\lim_V(A_V^F\Om,C^*B\Om)\end{equation}
We have already shown that the second term goes to zero. In the first
term the commutator becomes
\begin{equation}[A_V^F,C]=V^{-1/2}\cdot[A_{V_0},C]\end{equation}
and hence the first term goes also to zero. In case we assume only
$A\in\mcal{A}$ a further $L^1$-condition for the three-point function
is needed to arrive at the same result. As $\mcal{A}_0\Om$ is
assumed to be dense in $\mcal{H}$ and $\|A_V^F\|<\infty$ uniformly in $V$, we have
\begin{prop}$L^1$-clustering implies that
\begin{equation}A_V^F\to 0\quad\text{weakly on $\mcal{H}$}\quad,\quad\|A_V^F\Om\|<\infty\quad\text{uniformly in $V$}   \end{equation}
but $\|A_V^F\Om\|$ bounded away from zero in general. That is, $
A_V^F$ does not converge strongly to zero and, a fortiori, there is no
convergence in norm.
\end{prop}
This clearly shows that, in order to have non-trivial limit operators,
one has to leave the original Hilbert-space of microscopic observables
and has to define or construct an entirely new representation living
on a different state.
\subsection{A Smoothed Version of Fluctuation Operators}
Since we employ in the following so-called \tit{Fourier-methods} and
related calculational tools, it is advantageous to change to a
smoother version of fluctuation operators. As everybody knows, sharp
volume cut-offs are both a little bit artificial and technically
nasty, since they may sometimes lead to non-generic or spurious
effects. In other branches of rigorous statistical mechanics or
axiomatic quantum field theory volume integrations have therefore
frequently been emulated or implemented in a slightly different way
(see e.g. \cite{Requ2}).

Two choices have basically been in use with the second version having
much nicer properties in several respects as we will explain below.
Instead of integrating over a sharp volume, $V$, centered e.g. around
the coordinate origin, one integrates the shifted observable, $A(x)$,
over a smooth test function localized basically in $V$ but having smooth tails.\\[0.3cm]
Remark: As $V$ we choose in the following a ball centered at the
origin with radius $R$ and let $R$ go to infinity.
\begin{defi}Two admisssible families of test functions are the
  following ones: $f_R(x)\geq0$ smooth with
\begin{equation}f_R(x):=
\begin{cases}1 & \text{for $|x|\leq R$} \\0 & \text{for $|x|\geq R+h$}
\end{cases}\end{equation}
or
\begin{equation}f_R(x):= f(|x|/R)\quad\text{with}\quad f(s)=
\begin{cases}1 & \text{for $|x|\leq 1$} \\ 0 & \text{for $|x|\geq 2$}
\end{cases}
\end{equation}
\end{defi}
Note that the latter choice has much nicer behavior under Fourier
transform while working with the Fourier transforms of the former version or e.g. the
indicator function of the volume $V$ is quite cumbersome). On the
other hand, the latter version has tails which are also scaled.
\begin{lemma}
\begin{equation}\hat{f}_R(k)=const\cdot R^n\cdot \hat{f}(R\cdot
  k)\end{equation}
\end{lemma}
where here and in the following `` $const$'' denotes an (in this
context) irrelevant numerical factor which, a fortiori, may change in
the course of a calculation.  With the help of this smearing functions
we now define
\begin{defi}[Smooth Volume Integration] We redefine the fluctuation
  operators in the following way
\begin{equation} A_R^F:=R^{-n/2}\cdot\int A(x)\cdot
  f_R(x)d^nx\end{equation}
with $f_R$, unless otherwise stated, the family given in the second
example above (remember $\langle A\rangle:=0$).
\end{defi}
\section{The Limiting Case for Normal Fluctuations}
In order to arrive at a rigorous definition of fluctuation operators
in a certain limit state we will follow a line of arguments
 which may complement the treatment of Verbeure et al in
several respects. We will study directly the macroscopic limit of the
n-point functions with the help of certain \tit{momentum space
  methods}. As they are perhaps not so common in statistical physics
we will give the technical details below.
\subsection{Some Generalities}
Any n-point (correlation) function of the kind $\langle A_1(x_1)\cdots
A_n(x_n)\rangle$ with the $A_i(x_i)$ the translates of the observables
$A_i$ (which may also contain an implicit time variable $t_i$ which is
however kept fixed in the following) is written as $W(x_1,\ldots,x_n)$.
With the state $\Om$ being translation invariant we have
\begin{equation}W(x_1,\ldots,x_n)=W(x_1-x_2,\ldots,x_{n-1}-x_n)\end{equation}
To express cluster properties in a clear way, we introduce the
so-called \tit{truncated correlation functions} via the following
recursion relation:
\begin{equation}W(x_1,\ldots,x_n)=\sum_{part}\prod_{P_i}W^T(x_{i_1},\ldots,x_{i_k})\end{equation}
where the sum extends over all partitions of the set $\{1,\ldots,n\}$
into subsets $P_i$ with the elements in each subset ordered as
$i_1<i_2\ldots<i_k$. The first elements of the recursion are
\begin{equation}W(x)=W^T(x)=0\quad\text{in our case}\end{equation}
\begin{equation}W^T(x_1,x_2)=W(x_1,x_2)-W(x_1)W(x_2)\end{equation}
\begin{ob}In the truncated correlation functions the vacuum state, ground
  state or equilibrium state, $\Om$, has been eliminated in a {\em
    symmetric} way, so that we have, in a sense to be specified,
\begin{equation}W^T(x_1,\ldots,x_n)\to 0\quad\text{for}\quad
  \sup|x_i-x_j|\to\infty\end{equation}
\end{ob}

In this section we assume the following \tit{cluster property}
\begin{equation}W^T(x_1,\ldots,x_n)\in L^1\;\text{in the
    variables}\;\{x_1-x_2,\ldots,x_{n-1}-x_n\}\end{equation}

From the above we see that the original hierarchy of $n$-point
functions can be reconstructed from the new hierarchy of truncated
$n$-point functions, which have more transparent cluster properties.
The $L^1$-condition allows us to Fourier transform the
$W^T(x_1,\ldots,x_l)$ and we get from translation invariance:
\begin{multline}const\cdot\int\tilde{W}^T(p_1,\ldots,p_l)\cdot e^{-i\sum
    p_ix_i}\prod dp_i=\\W^T(x_1,\ldots,x_l)
=W^T(x_1-x_2,\ldots,x_{l-1}-x_l)\\= const\int
\hat{W}^T(p_1,p_1+p_2,\ldots,p_1+\cdots
p_{l-1})\cdot\delta(p_1+\cdots p_l)e^{-i\sum p_ix_i}\prod dp_i\\
= const\int \hat{W}^T(q_1,\ldots,q_{l-1})e^{-i\sum_{i=1}^{l-1}
  q_iy_i}\prod_{i=1}^{l-1} dq_i
\end{multline}
with
\begin{equation}\label{q}      y_i:= x_i-x_{i+1}\;,\;q_i=\sum_{j=1}^i p_j\quad i\leq
  (l-1)\end{equation}
The functional determinant $det(\partial q/\partial p)$ is one and we
can regard $\hat{W}^T$ either as a function of the $q_i$'s or the
$p_i$'s. We hence have
\begin{lemma} As a Fourier transform of a $L^1$-function\\ $\hat{W}^T(p_1,\ldots,p_{l-1})=\hat{W}^T(q_1,\ldots,q_{l-1})$
  is a continuous and bounded function which decreases at infinity in
  the $q$-variables.
\end{lemma}
\subsection{The $(k=0)$-Modes}
We now study the limit of truncated $l$-point functions with the entries being
fluctuation operators $A^F_R$, more precisely their Fourier
transforms, i.e.
\begin{multline}\langle A^F_R(1)\cdots A^F_R(l)\rangle^T=\\const\cdot
  R^{ln/2}\cdot\int \hat{f}(Rp_1)\cdots \hat{f}(-R[p_1+\cdots
 + p_{l-1}])\cdot \hat{W}^T(p_1,\ldots,p_{l-1})\prod dp_i\\
=const\cdot R^{ln/2}\cdot R^{-(l-1)n}\cdot\int \hat{f}(p_1')\cdots
\hat{f}(-[p_1'+\cdots+p_{l-1}'])\cdot
\hat{W}^T(p'_1/R,\ldots,p'_{l-1}/R)\prod dp'_i\end{multline}
$\hat{W}$ is continuous and bounded and the $\hat{f}$'s are of rapid
decrease. Hence we can perform the limit $R\to\infty$ under the
integral  and get
\begin{satz}The expression $\langle A^F_R(1)\cdots A^F_R(l)\rangle^T$ scales as $\sim R^{(2-l)n/2}$. This
 implies that for $l>2$ the above limit is zero, for $l=2$ the
  limit is a finite number bounded away from zero in general. In other
  words we have
\begin{equation}\lim_{R\to\infty} \langle
  A^F_R(1)\cdots A^F_R(l)\rangle^T=0\;\text{for}\;l>2\end{equation}
and
\begin{equation}\lim_{R\to\infty}\langle
  A^F_R(1)\cdots A^F_R(l)\rangle=
  \lim_{R\to\infty}\sum_{part}\prod_{\{ij\}}\langle A^F_R(i)A^F_R(j)\rangle\end{equation}
\end{satz}

The relation between the original microscopic system
$(\mcal{A},\omega)$ and the coarse-grained system of fluctuation
operators is a little bit subtle. Note that $\omega_F$, the limit state
to be constructed, can no longer be considered as a state or something
like that on the original algebra nor can the fluctuation operators be
considered as a representation of, say, $\mcal{A}$. One aspect of the
impending problems can perhaps best be seen by realizing that e.g.
\begin{equation}(A\cdot B)_V^F \neq A_V^F\cdot B_V^F\end{equation}
which pertains also in the limit. That is, in a sense to be defined,
we have
\begin{equation}(A\cdot B)^F\neq A^F\cdot B^F\end{equation}
the same holding in general for all the higher products. This is one
source of non-uniqueness as there is no invariant discrimination
between an observable regarded as a single object to be scaled and as
a product of other observables, where now each factor has to be scaled
separately. The appropriate point of view has to be a different one
(as has also been emphasized by Verbeure et al, cf e.g.
\cite{Verbeure1}, second ref.  p.540f and private communication).

The picture remains relatively clear for the intermediate
scales, $V<\infty$. We have a start system $(\mcal{A},\om)$, labelled
by, say, $V=0$. On every scale $V$ we have a new algebra,
$\mcal{A}^F_V$, (actually a subalgebra of $\mcal{A}$), generated by
the observables $A_V^F\,,\,A\in \mcal{A}$ (including arbitrary finite
products $(A_1\cdots A_n)^F_V$). If we prefer to consider this algebra
on scale $V$ as a new abstract algebra (i.e. forgetting about the
underlying finer algebra $\mcal{A}$), we get also a new,
coarse-grained state via the identification
\begin{equation}\om^F_V(\Pi A_V^{F,i}):=\om(\Pi A_V^{F,i})\end{equation}
(A related philosophy was expounded by Buchholz and Verch in
e.g. \cite{Buchholz} within the context of the algebraic analysis of
ultra-violet behavior in quantum field theory.)

The map
\begin{equation}R_V:\;\mcal{A}\to\mcal{A}^F_V\end{equation}
can be viewed as kind of a \tit{renormalization map}, which does
however \tit{not} preserve the algebraic structure (i.e.the algebras
are in general not \tit{isomorphic}). Furthermore one gets a ``\tit{new}''
dynamics on this algebra by defining
\begin{equation}\alpha^V_t(A_V^F):=(\alpha_t(A))^F_V\end{equation}
\begin{bem} In our context $\alpha_t$ is assumed to commute with the space
  translations or with a corresponding lattice version, that is, we
  have $\alpha_t(A^F_V)=(\alpha_t A)^F_V$. (Furthermore it may turn
  out to be reasonable to scale the time variable on the lhs also.)
\end{bem}

On the other hand, in order to construct the limit theory itself, one
can proceed in a slightly different direction.  The above limits of
n-point functions define a consistent hierarchy of new n-point
functions which then allow to define a \tit{new} limit system via the
so-called \tit{reconstruction theorem} (for a pendant in quantum field
theory see e.g. \cite{Wightman}). Put differently, we define limit
objects, $\{A_i^F\}$, the so-called fluctuation operators, which live
in a new Hilbert space built upon the new state, $\omega_F$, defined
by the limits:
\begin{equation}\label{limit}   \omega_F(A_1^F\cdots A_n^F):=
  \lim_{R\to\infty}\langle A^F_{1,R}\cdots
  A^F_{n,R}\rangle=\sum_{part}\prod_{\{ij\}}\omega_F(A^F_i\cdot
  A^F_j)\end{equation}

Note however that the so-called \tit{Gelfand-ideal}, $I_F$, is large,
that is, there are a lot of elements of $\mcal{A}$ which are mapped to
zero by this limit with
\begin{equation}I_F:=\{A\,;\,\omega_F((A^F)^*\cdot A^F)=0\}\end{equation}
This is of course typical for such kind of \tit{mean-values}, as
e.g. all space-translates of $A$ yield the same limit
element. Shifting one of the observables in the above $l$-point
functions by, say, $a_i$ yields an extra factor $e^{ip_ia_i}$ in the
Fourier transform which after the above coordinate tranformation goes
over into $e^{ip_i'/R\cdot a_i}$ which goes to one. Summing up we have
\begin{conclusion} With the help of equation (\ref{limit}) we construct a new limit
  system, consisting of the {\em algebra of fluctuation operators},
  $\mcal{A}_F$, and the limit state $\omega_F$. The well-known
  GNS-construction (see e.g. \cite{Bratteli1}) allows to construct the
  corresponding Hilbert-space representation with
\begin{equation}\omega_F(A_1^F\cdots A_n^F)=(\Om_F,A_1^F\cdots
  A_n^F\Om_F)\end{equation}
(where, by abuse of notation, we do not discriminate between operators
and their equivalence classes on the rhs).

As all the $n$-point functions decay into a product of $2$-point
functions all the commutators are $c$-numbers:
\begin{equation}[A^F,B^F]=\omega_F([A^F,B^F])\end{equation}
The system of fluctuation operators is a {\em quasi-free} system
(cf. \cite{ Bratteli2})
\end{conclusion}

Taking now self-adjoint elements one can, as in \cite{Verbeure1},
represent the new system as a representation of the $CCR$ over the real vector space of
s.a. operators. Our scalar product, induced by the hierarchy of
$n$-point functions, can be split in the following way.
\begin{equation}(A^F\Om_F,B^F\Om_F)=Re\;(\ldots)+i\,Im\;(\ldots)=:s_F(A^F,B^F)+(i/2)\sigma_F(A^F,B^F)\end{equation}
\begin{equation}\om_F([A^F,B^F])=\sigma_F(A^F,B^F)\end{equation}
where $\sigma_F$ defines a \tit{symplectic form}. The
\tit{Weyl-operators}, $e^{iA^F}$ with $A^F$ s.a., fulfill the
$CCR$-relations
\begin{equation}\om_F(e^{iA^F})=e^{-1/2s_F(A^F,A^F)}\end{equation}
\begin{equation} e^{iA^F}\cdot e^{iB^F}=e^{i(A^F+B^F)}\cdot e^{-i/2\sigma_F(A^F,B^F)}\end{equation}
In our context the first equation can e.g. be verified as
follows: Only the $2n$-point functions are different from
zero. On the lhs we hence have
\begin{equation}\label{exp}   \om_F(e^{iA^F})=\sum
  (-1)^n/(2n)!\cdot\om_F([A^F]^{2n})\end{equation}
It remains to count the number of
partitions of an $2n$-set into $2$-sets. This number is $(2n)!/2^n\cdot n!$.
In (\ref{exp}) we now get for $A^F$ s.a. on the rhs
\begin{equation}\sum_n 1/n!(-1/2\cdot
  \om_F(A^FA^F))^n=e^{-1/2s_F(A^FA^F)}\quad\Box\end{equation}

The above general cluster result of the limit $n$-point functions make
the study of the limit time evolution relatively straightforward. In a
first step it suffices to study the $2$-point functions. We define
the time evolution in the limit theory by
\begin{equation}\om_F(A^F(t')\cdot B^F(t)):=\lim\om(A_V^F(t')\cdot
  B_V^F(t))=\lim\om(A(t')_V^F\cdot B(t)_V^F)\end{equation}
 On the limiting GNS-Hilbert space constructed above we now get a
bounded \tit{sesquilinear form} $(x,y(t))$ which, by standard results,
yields a bounded operator $U^F(t)$ implementing the time
evolution. Here we use that the limit n-point functions are products
of $2$-point functions. Furthermore
we infer with the help of the above limit process that
\begin{equation}(U_t^Fx,U_t^Fy)=\om_F(\ldots)=\lim\om(\ldots)=(x,y)\end{equation}
In other words, we arrive at the following conclusion
\begin{satz}The preceding construction yields a strongly continuous
  unitary time evolution on the limiting GNS-Hilbert space.
\end{satz}

Another point worth to be mentioned (since it might perhaps be overlooked) is the question of the non-triviality
of the commutators
\begin{equation}[A^F,B^F]=\om_F([A^F,B^F])\end{equation}
In principle it could happen that all the expectation values on the
rhs vanish. In that case the limit algebra would be \tit{abelian} and
the fluctuations \tit{classical}. In a more general context
(cf. e.g. \cite{Buchholz}) this problem is more complicated. In our
situation this question can however be answered in a rather
straightforward way. We have
\begin{equation}\lim_V\om([A^F_V,B^F_V])=\lim_V\om([A_V,V^{-1}\cdot
  B_V])\end{equation}
For $A,B\in\mcal{A}_0$, i.e. local, the rhs equals
\begin{equation}\lim_V\om([A_V,B])\end{equation}
We know candidates which lead to a vanishing of the limit for all
$B\in\mcal{A}_0$. For $A$ chosen s.a. these are the generators of
\tit{conserved symmetries}, written
\begin{equation}Q:=\int A(x)d^nx\end{equation}
Usually they are assumed to commute with the time evolution, expressed
as $Q(t)=Q$, hence the above limit would also be zero on the full
quasi-local algebra.  This situation, more specifically the case of
\tit{spontaneous symmetry breaking (ssb)} and \tit{Goldstone
  phenomenon}, will be dealt with in more detail in section
\ref{Goldstone}. In any case, as conserved symmetries are usually not
so numerous, we may presume that, in the generic case, not all of
these commutators will be zero.

For $A,B$ not necessarily strictly local our above more general
formalism is useful. With
\begin{equation}\om(A(x)B)=F_{AB}(x)\quad,\quad\om(BA(x))=G_{AB}(x)\end{equation}
the vanishing of the commutator would imply:
\begin{multline}0=[A^F,B^F]=\lim_R R^n\cdot\int
  |\hat{f}(Rp)|^2(\hat{F}_{AB}(p)-\hat{G}_{AB}(p))d^np\\
=\lim_R\int |\hat{f}(p)|^2(\hat{F}_{AB}(p/R)-\hat{G}_{AB}(p/R))d^np\\
= (\hat{F}_{AB}(0)-\hat{G}_{AB}(0))\cdot\int|\hat{f}(p)|^2d^np
\end{multline}
by the theorem of dominated convergence (note that we are in the
$L^1$-situation). Hence we have the result
\begin{prop}
\begin{equation}[A^F,B^F]=0\Leftrightarrow
  \hat{F}_{AB}(0)=\hat{G}_{AB}(0)\end{equation}
that is
\begin{equation}\int F_{AB}(x)d^nx=\int G_{AB}(x)d^nx\end{equation}
or
\begin{equation}\int (\Om,[A(x),B]\Om)d^nx=0\end{equation}
which is the same result as in the strictly local case.
\end{prop}
\subsection{The $(k\neq 0)$-Modes}
Up to now only the $(k=0)$-modes of fluctuation operators, i.e.\\
$\lim_V V^{-n/2}\cdot\int_V A(x)d^nx$, have been studied. For various
reasons it is useful to have corresponding formulas at hand for
fluctuation observables containing a certain net-momentum. This
problem was studied by Verbeure et al in e.g. \cite{k-Mode} and the
results were applied in e.g. \cite{Boson} in the analysis of
\tit{Goldstone modes}. In the original (real-space) approach the
necessary calculations turned out to be quite involved and far from
being simple. This is another case in point to demonstrate the merits
of our Fourier space scaling methods.

Instead of the original scaling operators, $A_V^F$ or $A_R^F$, we now
study their $k\neq 0$-variants, $A_R^F(k)$. We begin with a technical lemma.
\begin{lemma}
\begin{equation}\hat{A}(k):=(2\pi)^{-n/2}\int e^{ikx}A(x)d^nx\end{equation}
is an operator-valued distribution
(We use the convention\\ $\hat{f}(k)=(2\pi)^{-n/2}\int e^{-ikx}f(x)d^nx$)
\end{lemma}
Remark: For a systematic use and proofs of such energy-momentum techniques in
quantum statistical mechanics we refer to e.g. \cite{Quasi} where also some
more mathematical background is
provided.\\[0.3cm]
Integrating now over $e^{iqx}\cdot f_R(x)$, we get the $q$-mode
fluctuation operators.
\begin{multline}A_R^F(q):=R^{-n/2}\int
  A(x)e^{iqx}f_R(x)d^nx=R^{n/2}\int \hat{A}(k+q)\hat{f}(Rk)d^nk\\
=R^{n/2}\int \hat{A}(k)\hat{f}(R(k-q))d^nk\end{multline}

We can now proceed in exactly the same way as above in the case of the
zero-mode analysis and calculate the truncated $l$-point functions
$\langle A_R^F(1,q_1)\cdots A_R^F(l,q_l)\rangle^T$ (where the indices
$1$ to $l$ label different observables). The only thing
that changes are the test functions, i.e. $f_R(x)\to e^{iq_kx}\cdot
f_R(x) $. We arrive at the conclusion:
\begin{satz}[$q$-Mode Fluctuation Operators]\hfill\\
In the case of $L^1$-clustering all truncated correlation functions
vanish for $l\geq 3$ and the $l$-point functions are again sums of
products of $2$-point functions. The concrete form of the
limit-$2$-point functions is given in formula (\ref{two}).
\end{satz}
If we calculate the limt-$2$-point functions
explicitly we get:
\begin{multline}\langle A_R^F(q_1)\cdot B_R^F(q_2)\rangle^T =\\
  R^n\int\langle \hat{A}(k_1+q_1)\hat{B}(k_2+q_2)\rangle^T\cdot
  \delta(k_1+q_1+k_2+q_2)\cdot \hat{f}(Rk_1)\hat{f}(Rk_2)dk_1dk_2\\
  =R^n\int\ \langle
  \hat{A}(k_1+q_1)\hat{B}(-(k_1+q_1))\rangle^T\cdot\hat{f}(Rk_1)\hat{f}(-R(k_1+q_1+q_2))dk_1\\
  =R^n\int\langle
  \hat{A}(k)\hat{B}(-k)\rangle^T\cdot\hat{f}(R(k-q_1))\hat{f}(-R(k+q_2))dk\end{multline}
With $k':=R(k-q_1)$ we arrive at
\begin{equation}\int \hat{W}^T(k'/R+q_1)\cdot
  \hat{f}(k')\hat{f}(-k'-R(q_1+q_2))dk'\end{equation}
By assumption $\hat{W}^T$ is in $L^1$, $\hat{f}$ is of rapid decrease,
so the limit can again be carried out under the integral and we have
\begin{ob}For $q_1+q_2\neq 0$ it holds
\begin{equation}\lim_R\langle A_R^F(q_1)\cdot B_R^F(q_2)\rangle^T
  =0\end{equation}
For $q=q_1=-q_2$ we get on the other side
\begin{equation}\label{two}\lim_R\langle A_R^F(q)\cdot
  B_R^F(-q)\rangle^T=\hat{W}^T(q)\cdot\int
  \hat{f}(k)\hat{f}(-k)dk\end{equation}
\end{ob}
In other words, the limit tests the spectral momentum of the two-point
function.
\section{The Case of $L^2$-Clustering}
Before we embark on an investigation of the situation in the regime
where phase transitions, vacuum degeneracy and/or spontaneous symmetry
breaking (ssb) prevail, we briefly address the case where the
clustering is weaker than $L^1$ but still $L^2$, say. Our above
Fourier-space approach can easily handle also this more singular situation.
We hence assume now that the truncated $l$-point functions cluster
only in the $L^2$-sense in the difference variables.

Now we cannot conclude that the Fourier transform. is bounded and
continuous, but we know it is again an $L^2$-function. We repeat the
first steps of the above calculation with, however, another
\tit{scaling exponent}, $\alpha$, which we leave open for the moment.
\begin{defi} In the general case we define fluctuation operators by
\begin{equation}A_R^F:=R^{-\alpha}\cdot\int
  A(x)f_R(x)d^nx\end{equation}
\end{defi}
We get
\begin{multline}\langle A_R^F(1)\cdots A_R^F(l)\rangle^T=\\const\cdot
  R^{l(n-\alpha)}\cdot\int
  \hat{f}(Rp_1)\cdots\hat{f}(-Rq_{l-1})\cdot
  \hat{W}^T(q_1,\ldots,q_{l-1})\prod dq_i
\end{multline}
where the $\{p_i\}$ are linear functions of the $\{q_i\}$ as described
above. We now apply the Cauchy-Schwartz inequality
\begin{multline}\label{L^2}|lhs|\leq const\cdot
  R^{l(n-\alpha)}\left[\int(\hat{f}(Rp_1)\cdots
    \hat{f}(-Rq_{l-1}))^2\prod
    dq_i\right]^{1/2}\\\cdot\left[\int(\hat{W}^T(q_1,\ldots,q_{l-1}))^2\prod dq_i\right]^{1/2}
\end{multline}
In the first integral on the rhs we make again a variable
transformation from $q_i$ to $q'_i:=Rq_i$, yielding an overall scaling
factor
\begin{equation}R^{l(n-\alpha)}\cdot R^{-(l-1)n/2}\end{equation}

We again want the limits of the $2$-point functions to be both finite and
non-trivial, i.e. different from zero in general.
\begin{prop} To make the rhs of (\ref{L^2}) finite in the limit for $l=2$ the {\em
    maximal} $\alpha$ to choose is
\begin{equation}3n-4\alpha=0\quad\text{i.e.}\quad
  \alpha=(3/4)n\end{equation}
For a general $l$ this leads to the scaling exponent $(n-(1/2)l\cdot n)/2$,
which is negative for $l\geq 3$. Hence, all higher truncated $l$-point
functions vanish in the limit.
\end{prop}
However, to guarantee that the result is really non-trivial, we have
to analyze the situation in more detail as the above estimate is only
an inequality. In the case of $L^1$-clustering $\alpha=n/2$ was
appropriate. The largest value which can occur in the $L^2$-case is
the above maximal $\alpha=(3/4)n$. If we want to avoid that the
$2$-point functions vanish in the limit we have to choose in the
$L^2$-case
\begin{equation}(1/2)n<\alpha\leq (3/4)n\end{equation}
depending on the concrete decay of the $2$-point functions in
configuration space. We see that, evidently, the situation is now less
canonical as compared to the $L^1$-case.\\[0.3cm]
Remark: A related situation (on a lattice) was analyzed by Verbeure et al in
\cite{Verbeure2}, where a clustering weaker than $L^1$ was considered
with, however, the additional input that the local algebras, sitting
at the points of the lattice, form a finite-dimensional
\tit{Lie-algebra}. In that case, suitable scaling exponents are chosen
to render the auto-correlation functions finite and non-vanishing,
while, on the other side, the finiteness of the limit $3$-point
functions has to be imposed as an extra assumption. Under this proviso
one gets the existence of a limit Lie-algebra, but nevertheless
results are only partial while perhaps, on the other side, being also
more interesting.\\[0.3cm]
We do not want to dwell too much on this point at the moment, as progress
seems to be to a certain extent model-dependent. Furthermore, we
develop a different approach in the last section which is able to cope
with any kind of poor cluster behavior.

If we want to guarantee the apriori existence or vanishing  of the
truncated $3$-point functions with the help of our above
$L^2$-estimate (\ref{L^2}), we have to restrict the chosen $\alpha$ in
the following way.
\begin{koro}If the appropriate $\alpha$ fulfills $\alpha>(2/3)n$, we get
  a negative scaling exponent for $l\geq 3$ as
\begin{equation}n-(1/3)ln\leq 0\quad\text{for}\quad l\geq
  3\end{equation}
For $\alpha=2/3$ the $3$-point functions are finite.
\end{koro}
\begin{bem}One would get corresponding relations for smaller $\alpha$
  but higher correlation functions, beginning from a certain order,
  $l_0(\alpha)$ say. On the other hand, one cannot guarantee the apriori
  existence of the $l$-point functions for $2<l<l_0(\alpha)$ as the
  general scaling relation reads for $l\geq l_0(\alpha)$:
\begin{equation}l(2\alpha-n)>n\quad\text{and}\quad
  \alpha>(1/2)n\end{equation}
and $\alpha$ being so chosen that the $2$-point functions are
non-trivial.
\end{bem}
\section{Spontaneous Symmetry Breaking (SSB) and the Goldstone Phenomenon}
\label{Goldstone}
\subsection{General Remarks}
Before we study fluctuation operators in the regime of vacuum--,
ground--,\\equilibrium--state degeneracy, we want to briefly comment, in
order to set the stage, on the (rigorous)
implementation of $ssb$ in the various areas with particular emphasis
on (quantum) statistical mechanics, i.e. condensed matter physics. As
this topic has however been much discussed in the past from various
points of views, we do not intend to give an exaustive commentary. We
only mention some earlier work being of relevance for our
argumentation and sketch the general framework.

We assume that our state, $\om$ or $\Om$, is (non-)invariant under
some automorphism group of $\mcal{A}_0$ or $\mcal{A}$. Furthermore,
and this is important (while frequently not clearly stated), we assume
the time evolution, $\alpha_t$, to commute with the automorphism
group, $\alpha_g$.
\begin{defi}$\alpha_g$ is called a symmetry group if
\begin{equation}\alpha_g\cdot\alpha_t=\alpha_t\cdot\alpha_g\end{equation}
\end{defi}
\begin{defi}If
\begin{equation}(\Om,\alpha_g(A)\Om)=(\Om,A\Om)\end{equation}
for all $A\in\mcal{A}$, the symmetry is called {\em conserved} and can
be implemented by a unitary group of operators in the representation
space
\begin{equation}\alpha_g(A)\;\to\;U(g)AU(g^{-1})\end{equation}
On the other side, if
\begin{equation}(\Om,\alpha_g(A)\Om)\neq(\Om,A\Om)\end{equation}
for some $A$, $A$ the {\em symmetry-breaking observable}, the symmetry
is called {\em spontaneously broken} since it still commutes with the
time evolution (i.e. formally: with the Hamiltonian, modulo {\em
  boundary terms} due to {\em long-range correlations}).
\end{defi}

In most cases the (continuous) symmetry group derives from a clearly
identifiable \tit{generator} (we restrict ourselves, for convenience,
to one-parameter groups) which is built from a local operator density,
i.e.
\begin{equation}U(s)=e^{isQ}\quad,Q(t)=\int
  q(x,t)d^nx\;,\;Q(t)=Q(0):=Q\end{equation}
Note that there are a lot of technical subtleties lurking behind these
operator identities, all of which we cannot mention in the following
(for more details and references see e.g. \cite{ssb}. A nice review is
\cite{Wre} where many of the widely scattered results have been compiled
).
\begin{bem}In many situations the generator density is the zero-component of a
  {\em conserved current}. Formally the conservation law encodes the
  time-independence of the global charge, $Q$. Furthermore, for
  convenience, we assume the symmetry to commute with the space
  translations, i.e. $U(x)QU(-x)=Q$. This is in fact frequently the
  case and simplifies certain calculations.
\end{bem}
The most crucial consequence is that in case the symmetry is
spontaneously broken some of the above relations do only hold in a
formal or algebraic sense. More specifically:
\begin{satz}If $\alpha_g$ is spontaneously broken the global generator $Q$ does only exist
  in a formal sense as a limit
\begin{equation}Q=\lim_V Q_V\quad,\quad Q_V:=\int_V
  q(x)d^nx\end{equation}
We have
\begin{equation}ssb\Leftrightarrow\lim_V(\Om,[Q_V,A]\Om)\neq 0
  \label{Q}\end{equation}
for some $A\in\mcal{A}$ and $Q$ is in that case only definable as a {\em nasty}
operator (see below).
\end{satz}
In the following we will take (\ref{Q}) as the defining relation of
$ssb$ (the technical details of the various statements can be found in
the literature, mentioned above).

The notion of $ssb$ is closely connected with another phenomenon, the
so-called \tit{Goldstone-phenomenon}. While there exists a clear
picture in, say, \tit{relativistic quantum field theory}, the
corresponding picture is a little bit blurred in the non-relativistic
regime. In the relativistic context we have sharp \tit{zero-mass
  Goldstone-modes}, i.e. true particles due to relativistic
covariance. On the other hand, in e.g. condensed matter physics or
statistical mechanics the situation is less generic. In general we do
no longer have sharp excitation modes; we have rather to expect
excitation modes having a \tit{finite lifetime} for momentum different
from zero but becoming infinitely sharply peaked for momentum $k\to
0$. The proper view is it to analyze these excitation branches in the
\tit{full} Fourier-space of \tit{energy-momentum} as has e.g. been
done in ref. four of \cite{ssb} and earlier in the author's doctoral
thesis, the principal object being the spectral-resolution of the
$2$-point correlation functions (in a neighborhood of $(E,k)=(0,0)$).
$SSB$ or the Goldstone phenomenon manifests itself in this quantity by
a singular contribution in the spectral measure. One should mention at
this place the work of Bros and Buchholz (see e.g. \cite{Bros}) about
quantum field theory in temperature (i.e. \tit{KMS}-) states. In this
particlar context the residual causality and locality properties of
the underlying relativistic theory lead to a, in some respects, more
generic behavior as compared to the ordinary non-relativistic
condensed matter regime.

In the non-relativistic regime it turns out that the concrete
structure of the Goldstone mode depends usually on the details of the
microscopic interactions (that means both the so-called
energy-momentum dispersion-law which can be, to give an example,
quadratic or linear near $k=0$ in the case of magnons or phonons, say,
and the $k$-dependent width of the branch). This led to the desire to
characterize the presence of a Goldstone phenomenon by a simple (if
qualitative) property. Sometimes one finds in the literature the
saying that the Goldstone phenomenon consists of the vanishing of a
\tit{mass-gap} above the ground state. But this statement is in some
sense frequently empty. From \cite{Swieca} we know e.g. that a
\tit{short-ranged} \tit{Galilei-covariant} theory, with a
non-vanishing particle density, cannot have a mass-gap due to
\tit{phonon-excitations} which signal the trivial breaking of the
Galilei-boosts. Furthermore, in most cases KMS-Hamiltonians have as
spectrum the whole real line.
\begin{bem}Models like the famous BCS-model (having a gap) are no case
  in point as they are implicitly breaking Galilei-invariance as do
  all such {\em mean-field-models}. This becomes apparent when
  analyzing the interaction part of the corresponding Hamiltonian. The
  complete fermion- or boson-liquid is, on the other side, again
  Galilei-invariant, hence has no mass-gap, but may, of course, still
  display e.g. {\em superfluidity}.
\end{bem}
In the next subsection we will provide a, as we think, more satisfying
and completely general characterization of the Goldstone phenomenon
which is independent of the details of the model under discussion.
\subsection{Some Rigorous Results for the Symmetry Generator in the
  Presence of SSB}
After the above introductory remarks we want to prove a couple of
rigorous results which characterize to some extent the presence of
$ssb$ in the (non-)relativistic regime. The main observation is that
the symmetry generator is no longer defined as a nice operator in the
representation (Hilbert- or $GNS$-) space when $ssb$ is present and
that this, at first glance, mathematical result encodes some
interesting physics.

Let us work, for simplicity, in the context of \tit{temperature
  states}. This has the advantage that $\Om$ is separating, i.e.
\begin{equation}A\Om=B\Om\Rightarrow A=B\end{equation}
The first task is to give $Q:=\lim_V Q_V$ a rigorous meaning. The
standard procedure (see the above mentioned literature) is to define
$Q$ via:
\begin{equation}QA\Om:=\lim_V [Q_V,A]\Om\quad,\quad
  Q\Om:=0\end{equation}
for e.g. $A\in\mcal{A}_0$. For $V$ sufficiently large, the commutator
on the rhs becomes independent of $V$, hence there is a chance to get
a well-defined $Q$ (at least on a dense set of vectors) as on the lhs
we have by \tit{separability}
\begin{equation}A\Om=B\Om\Rightarrow A=B\Rightarrow
  [Q_V,A-B]=0\end{equation}
For $A\in\mcal{A}$ one has to employ cluster properties.
\begin{ob}We have already seen above that, while such a $Q$ may
  exist, the corresponding $\|Q_V\Om\|$ will nevertheless diverge for
  $V\to\R^n$! This shows that the connection between the global
  generator and its local approximations is not that simple. The best
  one can usually expect, even in the case of \tit{symmetry
    conservation}, is a weak convergence on a dense set
\begin{equation}(B\Om,QA\Om)=\lim_V(B\Om,Q_V\Om)\end{equation}
but, due to the above divergence of $\|Q_V\Om\|$, we cannot even have
weak convergence on the full Hilbert-space. (For more details see the
above cited literature; in particular \cite{ssb}, third ref., where
the various possibilities in the respective fields have been compared)
\end{ob}

We see from the above that $Q$ can be defined as a densely defined
operator but usually we want to have more. A conserved continuous symmetry
is given by a s.a. generator. Let us see under what conditions the
above $Q$ is at least \tit{symmetric} provided that the $Q_V$ are
symmetric. We assume the symmetry to be conserved, i.e.
\begin{equation}\lim_V(\Om,[Q_V,A]\Om)=0\quad\text{for all}\quad
  A\in\mcal{A}\end{equation}
We then have
\begin{multline}(B\Om,QA\Om)=\lim_V(B\Om,[Q_V,A]\Om)\\
=\lim_V\left(([Q_V,B]\Om,A\Om)+(Q_v\Om,B^*A\Om)-(A^*B\Om,Q_V\Om)\right)
\end{multline}
\begin{conclusion}$Q$ is symmetric if $\lim_V(A\Om,Q_V\Om)=0$ for all
  $A\in\mcal{A}_0$. Under the same proviso it follows
\begin{equation}(B\Om,QA\Om)=\lim_V(B\Om,Q_V\Om)\end{equation}
\end{conclusion}
What is the situation if the symmetry is spontaneously broken? For
convenience we replace again the sharp volume-integration by our
smooth one, i.e.
\begin{equation}Q_V\to Q_R:=\int q(x)f_R(x)d^nx\end{equation}
We know that there exists a symmetry-breaking observable $A$ s.t.
\begin{equation}\lim_R(\Om,[Q_R,A]\Om)\neq 0\Rightarrow
  QA\Om=\lim_R[Q_R,A]\Om\neq 0\end{equation}
Due to the assumed translation invariance, i.e.
\begin{equation}U(a)QU(-a)=Q\quad\text{or, what is the same,}\quad U(a)q(x)U(-a)=q(x+a)\end{equation}
 we have
\begin{equation}(\Om,QA\Om)=(\Om,Q\cdot V^{-1}A_V\Om)\end{equation}
and
\begin{equation}Q\cdot V^{-1}A_V\Om=V^{-1}\int_V
  U(x)d^nx\cdot QA\Om\end{equation}
$U(x)$ the unitary representation of the translations.\\[0.3cm]
Remark: As a result of a discussion with Detlev Buchholz, following a
seminar talk about the paper, we will give a technically more detailed proof of
the above statement in the appendix at the end of the paper. This
seems to be advisable since, as we are showing below, the global
operator, $Q$, turns out to be non-closable, which will make certain
limit-manipulations more cumbersome.\\[0.3cm]
\begin{lemma}
\begin{equation}s-\lim_V V^{-1}\int_V U(x)d^nx=P_{\Om}\end{equation}
$P_{\Om}$ the projector on the (in our case) unique vacuum-,ground-,
equilibrium-state.
\end{lemma}
Proof: The result is well-known (see e.g. \cite{Ruelle}). We give
however a very short and slightly different proof using our smooth volume integration.
With $V_R:=\int f_R(x)d^nx$, a \tit{spectral resolution} yields
\begin{equation}V_R^{-1}\cdot\int U(x)f_R(x)d^nx=const\cdot\left(\int
  f(x)d^nx\right)^{-1}\cdot\int\hat{f}(Rp)dE_p\end{equation}
Applied to a vector $\psi$ we can now employ Lebesgue's theorem of
dominated convergence and get
\begin{equation}\lim_V
V^{-1}\int
U(x)d^nx\cdot\psi=(\hat{f}(0))^{-1}\cdot\hat{f}(0)P_{\Om}\psi=P_{\Om}\psi\qquad\Box
\end{equation}
This yields
\begin{equation}\label{Gold}0\neq P_{\Om}QA\Om=\lim_V Q\cdot
  V^{-1}A_V\Om\end{equation}
On the other hand
\begin{equation}\lim_V\|V^{-1}A_V\Om\|=\|P_{\Om}A\Om\|=0\end{equation}
by an analogous reasoning (note that we assumed $(\Om,A\Om)=0$).

We have now a sequence of vectors, $V^{-1}A_V\Om$, converging to zero
in norm while $Q\cdot V^{-1}A_V\Om$ converges to $P_{\Om}QA\Om\neq 0$.
Summing up what we have shown we arrive at the following conclusion:
\begin{conclusion}[Goldstone Theorem]If we have $ssb$ and a
  separating vector, $\Om$, (representing the ground or temperature
  state), $Q$ can still be defined as an operator which is however not
  closable, hence, a fortiori, not symmetric (note that symmetric
  operators are closable). This abstract result has as a
  practical consequence the physical property exhibited in the
  preceding formulas. They express the content of the Goldstone
  phenomenon in the most general and model independent way.  We infer
  that $Q$ induces transitions from a singular part of the continuous
  spectrum, passing through $(E,p)=(0,0)$, to the extremal invariant
  state $\Om$. On the other side, a conserved symmetry implies
\begin{equation}Q\Om=0\;,\; P_{\Om}[Q,A]\Om=0\Rightarrow
  P_{\Om}QA\Om=0\end{equation}
\end{conclusion}

We show now that the above result really contains the original
Goldstone phenomenon. Let us e.g. assume that we have the above result
and, on the other side, a gap in the energy spectrum above the state
$\Om$. We emphasized above that an important ingredient of the notion
of $ssb$ is the time independence of, say, the above expression. We
employ again the spectral resolution of operators with respect to
energy-momentum. We hence have
\begin{equation}0\neq c=P_{\Om}Q\int
  \hat{A}(k,E)e^{-itE}dkdE\Om\end{equation}
with $c$ being independent of $t$. We choose a real testfunction
$g(t)$ with $\int g(t)dt=1$. This yields
\begin{equation}0\neq c=P_{\Om}Q\int A(t)\cdot g(t)dt\Om=
 P_{\Om}Q\int\hat{A}(E)\hat{g}(E)dE\Om\end{equation}
If there is a gap above zero we may choose the support of $\hat{g}$
so that
\begin{equation}supp(\hat{g})\cap supp(spec(H))=0\end{equation}
Since, by assumption, $P_{\Om}$ has been extracted in the
energy-support of $A$, we get the result $c=0$, that is, no symmetry breaking.
But we can infer more about the nature of the energy-momentum spectrum
near $(0,0)$. We see that $P_{\Om}QA(g(t))\Om$ depends only on the
value of $\hat{g}(E)$ in $E=0$, which is one in our case, but not on
the shape of $g$. Inspecting equation (\ref{Gold}) we can infer the
following: The Fourier transform of the rhs contracts around $k=0$ in
the limit $V\to\infty$. On the other side we learned that in the limit
both sides have their energy support concentrated in $E=0$. The lhs
shows that the limit vector is parallel to $\Omega$. Whereas we do not
want to go into the partly intricate details of the limiting processes
of \tit{non-closable operators} (note that it is e.g. dangerous to use
the adjoint, $Q^*$, in the reasoning as it is not densely defined),
the latter part of the above theorem should now be obvious.

This sharp excitation around $(E,k)=(0,0)$ extends
into the full energy-momentum plane in form of a (usually) smeared
excitation branch (having a finite $k$-dependent life-time). For the
regime of temperature states the situation was analyzed in some detail
in the fourth reference of \cite{ssb} and already in the authors
doctoral thesis. We see from the above that a similar situation
prevails in the more general case of a separable $\Om$ and,
analogously, for ground-state models where $Q$ can be defined in the
above way. Even if the above $Q$ is not definable as a non-closable
limit operator we arrive at a similar result by exploiting the
limit-expectation values instead of the strong vector- or operator
limits, but we do not want to dwell more into the corresponding
details in this paper which deals with a different topic.

\section{The Canonical (Goldstone) Pair in the Presence of $SSB$}
As far as we can see, the notion of a \tit{canonical Goldstone pair}
was introduced by Verbeure et al. in \cite{Boson}. In the following
section we want to prove only a few general (model-independent)
results, whereas much more could be shown by combining the framework,
developed above, with the techniques mentioned in the preceding section.

We remarked above that $ssb$ is characterized by the non-vanishing
(but time-independence) of the following commutator limit
\begin{equation} \label{sym}  0\neq c=\lim_V(\Om,[Q_V,A(t)]\Om)\end{equation}
To fix the notation: usually a pure phase is characterized by the
non-vanishing of a so-called \tit{order parameter} in the presence of
$ssb$. This is an observable, $B$ say, with
\begin{equation}
(\Om,B\Om)=\begin{cases}
c\neq 0 & \text{in the broken phase}\\
0 & \text{in the conserved phase (above $T_c$, say)}
\end{cases}
\end{equation}
From (\ref{sym}) we see that as order parameter we have to choose
\begin{equation}B:=\lim_V [Q_V,A]\end{equation}
while $A$ is the symmetry breaking observable.
\begin{bsp}In the Heisenberg-ferromagnet with spontaneous
  magnetization in, say, the $z$-direction the order parameter is
  $S_z$ or $\langle S_z\rangle$. As generator of the broken symmetry
  one may take $\sum S_x$ and as symmetry breaking obsrvable
  e.g. $S_y$.
\end{bsp}
We have seen that we can write
\begin{equation}0\neq c=\lim_V (\Om,[Q_V,A]\Om)=\lim_V
  (\Om,[Q_V,V^{-1}A_V]\Om)=\lim_R (\Om,[Q_R,V_R^{-1}A_R]\Om)\end{equation}
where
\begin{equation}Q_R:=\int q(x)f_R(x)d^nx\;,\;A_R:=\int_{S_R}
  A(x)d^nx\end{equation}
with $V_R$ the volume of the sphere, $S_R$, with radius $R$.

We can now split the scaling exponent among the two observables (the
volume of the unit sphere being absorbed in the constant).
\begin{equation}0\neq const = \lim_R (\Om,[R^{-\alpha}Q_R,R^{-(n-\alpha)}A_R]\Om)\end{equation}
This form of scaling may yield something reasonable if the scaling
exponents can be so adjusted that also
\begin{equation}(\Om,R^{-\alpha}Q_RR^{-\alpha}Q\Om)\;\text{and}\;(\Om,R^{-(n-\alpha)}AR^{-(n-\alpha)}A\Om)\end{equation}
remain finite in this limit.

In general it does not seem to be easy to get both rigorous and general
estimates on the scaling behavior of these quantities. Fortunately, in
the case of temperature (KMS) states, such estimates are available. In
\cite{Perez} to \cite{Requardt-KMS} the following special
(\tit{real-space}-) version of the \tit{Bogoliubov-Inequality} has
been proved and employed for the observables $Q_R$ and $V_R^{-1}A_R$:
\begin{equation}|\langle[Q_R,V_R^{-1}A_R]\rangle|^2\leq\langle
  V_R^{-1}A_RV_R^{-1}A_R\rangle\cdot\langle[Q_R,[Q_R,H]]\rangle\end{equation}
The delicate term is the double commutator on the rhs. If $Q$ is
spontaneously broken, boundary terms will survive in the commutator of
$Q_R$ and the Hamiltonian, $H$, when taking the limit $R\to\infty$,
while in a formal sense they
commute. The double commutator saves us two powers of $R$, so to
say. That is we arrive after some cumbersome manipulations at
\begin{equation}\langle[Q_R,[Q_R,H]]\rangle\sim
  R^{(n-2)}\;\text{for}\;R\to\infty\end{equation}
hence
\begin{equation}\langle V_R^{-1}A_RV_R^{-1}A_R\rangle\gtrsim
  R^{(2-n)}\;\text{for}\;R\to\infty\end{equation}
as the limit on the lhs is a constant different from zero in the case
of $ssb$.
\begin{satz}For temperature states we have  for the symmetry
  breaking observable
\begin{equation}\langle A_RA_R\rangle\gtrsim
  R^{(n+2)}\end{equation}
That is, compared with the ordinary, normal scaling behavior ($\sim
R^n$), the divergence is worse. From this one infers the following decay of
the two-point correlation function itself:
\begin{equation}|\langle A(x)A\rangle|\gtrsim R^{(n-2)}\end{equation}
\end{satz}
\vspace{0.3cm}
Putting all the pieces together we now have to make the following
identification:
\begin{equation}n-\alpha\geq (n+2)/2\;\Rightarrow\;\alpha\leq
  (n-2)/2\end{equation}
in order that the limit commutator is non-trivial, i.e. non-classical.
On the other hand, the divergence behavior of $\langle Q_RQ_R\rangle$
can frequently be inferred either from covariance properties (as in
relativistic quantum field theory; see e.g. the third reference in
\cite{Requ2}) or from an analysis of the spectral behavior in concrete
(non-relativistic) models. Summing up we have:
\begin{conclusion}[Canonical Pair] For a covariant four-current in
  relativistic quantum field theory the two-point function in Fourier
  space contains a prefactor $\sim p^2$ which yields (after some
  calculations) an $\alpha=1/2$ (for space dimension, $n=3$). On the
  other side, if we do not have such nice covariance properties the
  divergence of $\langle Q_RQ_R\rangle$ is generically much worser
  than $\sim R$ (in three dimensions). This holds, in particular, for
  the above temperature states. It follows that for temperature states
  we cannot find a critical exponent $\alpha$ so that both the
  auto-correlations remain finite in the limit and the commutator
  non-trivial. That is, for temperature states the limit fluctuations
  are classical (an observation already made by Verbeure et al for
  special models, see e.g. \cite{Boson}).
\end{conclusion}

The situation seems to be less generic for ground state models, i.e.
the temperature-zero case. For one, we do not automatically have an a
priori estimate as in the above conclusion, from which we can infer
that it is the autocorrelation of $A_R$ which is ill-behaved. For
another, in temperature states, as was shown in e.g. the fourth
reference of \cite{ssb} by the author, the spectral weight has to
become infinite along the Goldstone excitation branch in a specific
way (which is governed by the dispersion law of the Goldstone mode)
for energy-momentum approaching zero. This sort of singularity is
mainly responsible for the poor decay of the respective
auto-correlation function. This phenomenon may be absent in the
case of ground states as has also been shown for certain Bose-gas
models in \cite{Boson} where some of these questions have been dealt
with in greater detail. Note in particular that a variety of
aspects may depend on the precise shape of the Goldstone mode near
energy-momentum equal to $(0,0)$ as was shown in the above mentioned
paper of the author or in the unpublished doctoral thesis.

On the other side, there has been some interesting work of Pitaevskii
and Stringari (see e.g. \cite{Stringari}), who showed that variants of
the \tit{uncertainty principle} may lead to non-trivial results in
certain cases for ground state systems if one can exploit and control
certain additional \tit{sum rules}.
\begin{bem}Note that the ordinary uncertainty principle (for
  e.g. hermitean operators and ignoring possible domain questions)
  reads
\begin{equation}1/4\cdot|\langle [A,B]\rangle|^2\leq\langle
  AA\rangle\cdot\langle BB \rangle\end{equation}
\end{bem}
One sees that instead of the double commutator of the local symmetry
generator and the hamiltonian now a term like $\langle Q_RQ_R \rangle$
occurs. While we have an a priori estimate of the large-R-behavior of
the double commutator, the behavior of $\langle Q_RQ_R \rangle$ is
probably less generic (in particular in the ground state situation)
and we need some extra information of the kind mentioned above.

\section{The Case of SSB or Very Poor Decay of Correlations}
In the preceding sections we studied the case of $L^1$- or
$L^2$-clustering. In this last section we want to briefly show how we
can proceed in the case of extremely poor clustering. We want
however, for the sake of brevity and in order to better illustrate the
method, to concentrate on the simpler case of a uniformly poor decay
of all the correlation functions we are discussing. This is of course
not always the case but the scheme can be easily generalized (we discuss
this topic in more detail in \cite{renorm}, where we treat this
question in the context of the renormalisation group analysis).

We hence assume that the truncated $l$-point functions cluster weaker
than $L^2$ or $L^1$, say, in the difference variables,
$y_i:=x_{i+1}-x_i$, (see section 3.1). The following reasoning works
both in the case of non-$L^1$ or non-$L^2$ clustering. In the latter
case one would again use the \tit{Cauchy-Schwarz-inequality} (as in
section 3.2). To illustrate the
method we choose the non-$L^1$ procedure.\\[0.3cm]
So let us assume
\begin{equation}W^T(y_1,\ldots,y_{l-1})\not\in L^1\end{equation}

For each $l$ we assume the existence of a weight factor with a
suitable exponent, $\alpha_l\in \R$:
\begin{equation}P_l(y):=(1+\sum
  y_i^2)^{\alpha_l/2}\end{equation}
so that
\begin{equation}F(y):=P_l(y)^{-1}\cdot W^T(y)\in L^1\quad\text{for}\quad\alpha_l>\alpha_l^{inf}   \end{equation}
On the other side, we define the fluctuation operators with the
exponent $\gamma$, which will be adjusted later
\begin{equation}A_R^F:=R^{-\gamma}\cdot A_R\end{equation}
It follows
\begin{equation}W^T(y)=P_l(y)\cdot F(y)\end{equation}
with $F(y)$ an  (in general, $l$-dependent) $L^1$-function.

For the limit correlation functions we then get
\begin{equation}\langle A_R^F(1)\cdots A_R^F(l)\rangle^T=R^{ln}\cdot
  R^{-l\gamma}\cdot\int \hat{F}(q)\cdot \hat{P}_l(q)
 \left[ \hat{f}(Rp_1)\cdots \hat{f}(-Rq_{l-1})\right]\prod dq_i\end{equation}
(cf. section 3.1)
\begin{bem} We write the Fourier transform of $P_l(y)$ formally as
\begin{equation}\hat{P}_l(q)=(1+\sum
  D_{q_i}^2)^{\alpha_l/2}\end{equation}
(with $D_{q_i}$ the partial derivations). For non-integer $\alpha_l/2$ this is a \tit{pseudo-differential
  operator}. At the moment, for the sake of brevity, we do not want to say more about the
corresponding mathematical framework (see \cite{renorm} for a complete
discussion). What we in fact only need are the scaling
properties of the expression. If one wants to be careful one may
equally well take the explicit expression for the Fourier transform of
the above polynomial in the $y$-coordinates applied to the product of
the $f_R$'s and exploit its scaling properties.
\end{bem}
In any case, we get (with this proviso) and the usual variable
transformation $p_i':=Rp_i$:
\begin{multline}\langle A_R^F(1)\cdots
  A_R^F(l)\rangle^T=\\R^{ln-l\gamma-(l-1)n+\alpha_l}\cdot\int
  \hat{F}(q'/R)\cdot(R^{-2}+\sum  D_{q'_i}^2)^{\alpha_l/2} \left[ \hat{f}(p_1')\cdots
  \hat{f}(-q_{l-1}')\right]\prod dq_i'\end{multline}

Again only the explicit scaling prefactor matters in the limit
$R\to\infty$. (Note that for non-minimal $\alpha_l$ we may have
$\hat{F}(0)=0$. Technical intricacies like this one will be discussed
at length in \cite{renorm}). To get a finite result for \tit{all}
correlation functions we have to adjust the scaling parameter,
$\gamma$, so that the exponents vanish or are negative. We
choose $\alpha_2$ for $l=2$ so that the limit two-point function is
finite and non-vanishing. That is:
\begin{equation}n-2\gamma+\alpha_2=0\rightarrow \gamma=(n+\alpha_2)/2 \end{equation}
Inserting this $\gamma$ in the
general expression for $l\geq 3$, we conclude that the scaling
prefactor is finite in the limit provided that
\begin{equation}\alpha_l\le l\gamma-n=((l-1)n+l\alpha_2)/2 \end{equation}
with $\gamma$ fixed by the two-point function. For $\alpha_l<
l\gamma-n$ we can even conclude that all(!) higher limit correlation
functions vanish and that the resulting theory is (quasi-)free. The
latter would, for example, be the case if
\begin{equation}\alpha_l\le (l-1)\cdot\alpha_2\end{equation}
holds, since we then have (with $\alpha_2<n$):
\begin{equation}\alpha_l\le
  (l-1)\cdot\alpha_2<(l-1/2)\alpha_2=(2l-1)\cdot\alpha_2/2<
  ((l-1)n+l\alpha_2)/2\end{equation}
but nothing can be concluded in general for, say, $\alpha_l=l\cdot \alpha_2$.

We see that it is of tantamount importance to better understand the
assymptotic behavior of truncated $l$-point functions and, in
particular, the rate of decay as a function of $l$. We address this
topic in more detail in \cite{renorm}.

\section*{Appendix}
The rigorous implementation of the formula
\begin{equation}U(a)q(x)U(-a)=q(x+a)\end{equation}
is
\begin{multline}U(a)Q_RU(-a)=U(a)\int q(x)f_R(x)d^nxU(-a)=\int
  q(x+a)f_R(x)d^nx\\
=\int q(y)f_R(y-a)d^ny=:Q_R(a)
\end{multline}
The first question is: how does the global $Q$ behave under
translations? To answer this question we have to take recourse to the
definition of the global $Q$ as a limit of local operations. We have
\begin{equation}U(a)QA\Omega=U(a)\lim_R[Q_R,A]\Omega=\lim_R[Q_R(a),A(a)]\Omega\end{equation}
since it holds
\begin{equation}\lim_n U(a)\psi_n=U(a)\lim_n\psi_n\end{equation}
as $U(a)$ is bounded.
If $A$ is local we have for sufficiently large $R$ (and hence, in the limit):
\begin{equation}\lim_R[(Q_R(a)-Q_R(0)),A(a)]=0\end{equation}
We hence arive at
\begin{equation}U(a)QA\Omega=\lim_R[Q_R,A(a)]\Omega=QA(a)\Omega=QU(a)A\Omega\end{equation}
\begin{lemma}On the dense set $\mcal{A}_0\Omega$, $Q$ commutes with the
  translations.
\end{lemma}

In a next step we have to analyse the action of $Q$ on integrals or
averages like $\int_V U(x)AU(-x)d^nx\Omega$. More specifically, we
want to show that $Q$ commutes, so to speak, with the operation of
\tit{integration}. We have
\begin{equation}Q\cdot\int_V A(x)d^nx\Omega:=\lim_R[Q_R,\int_V
  A(x)d^nx]\Omega\end{equation}
We approximate the integral by a sum, that is:
\begin{equation}\int_V A(x)d^nx\psi:=\lim_i\sum_i d^nx_i\cdot
  A(x_i)\psi\end{equation}
and get (as the $Q_R$ are assumed to be nice, that is, closed
operators)
\begin{equation}[Q_R,\int_V A(x)d^nx]\Omega=\lim_i[Q_R,\sum_i
  d^nx_i\cdot A(x_i)]\Omega=\lim_i\sum_id^nx_i\cdot
  U(x_i)[Q_R(-x_i),A]\Omega\end{equation}
We again choose $R$ so large that
\begin{equation}[Q_R(-x),A]=[Q_R,A]\quad\text{for all $x\in
    V$}\end{equation}
which leads to
\begin{equation}[Q_R,\int_V A(x)d^nx]\Omega=\lim_i\sum_i d^nx_i\cdot
  U(x_i)[Q_R,A]\Omega=\int_V U(x)d^nx\cdot[Q_R,A]\Omega\end{equation}
Taking now the limit $R\to\infty$, we get
\begin{lemma}
\begin{equation}Q\int_V A(x)d^nx\Omega=\int_V U(x)d^nx\cdot
  QA\Omega\end{equation}
\end{lemma}
This shows, that our manipulations can be justified.

\end{document}